\documentclass[preprint,superscriptaddress,prb]{revtex4}

\usepackage{graphicx}

\begin{document}

\title{Magneto-exciton limit of quantum Hall breakdown in graphene}

\author{A. Schmitt}\email{aurelien.schmitt@phys.ens.fr}
\affiliation{Laboratoire de Physique de l'Ecole normale sup\'erieure, ENS, Universit\'e
PSL, CNRS, Sorbonne Universit\'e, Universit\'e de Paris, 24 rue Lhomond, 75005 Paris, France}
\author{M. Rosticher}
\affiliation{Laboratoire de Physique de l'Ecole normale sup\'erieure, ENS, Universit\'e
PSL, CNRS, Sorbonne Universit\'e, Universit\'e de Paris, 24 rue Lhomond, 75005 Paris, France}
\author{T. Taniguchi}
\affiliation{Advanced Materials Laboratory, National Institute for Materials Science, Tsukuba,
Ibaraki 305-0047,  Japan}
\author{K. Watanabe}
\affiliation{Advanced Materials Laboratory, National Institute for Materials Science, Tsukuba,
Ibaraki 305-0047, Japan}
\author{J.M. Berroir}
\affiliation{Laboratoire de Physique de l'Ecole normale sup\'erieure, ENS, Universit\'e
PSL, CNRS, Sorbonne Universit\'e, Universit\'e de Paris, 24 rue Lhomond, 75005 Paris, France}
\author{G. M\'enard}
\affiliation{Laboratoire de Physique de l'Ecole normale sup\'erieure, ENS, Universit\'e
PSL, CNRS, Sorbonne Universit\'e, Universit\'e de Paris, 24 rue Lhomond, 75005 Paris, France}
\author{C. Voisin}
\affiliation{Laboratoire de Physique de l'Ecole normale sup\'erieure, ENS, Universit\'e
PSL, CNRS, Sorbonne Universit\'e, Universit\'e de Paris, 24 rue Lhomond, 75005 Paris, France}
\author{G. F\`eve}
\affiliation{Laboratoire de Physique de l'Ecole normale sup\'erieure, ENS, Universit\'e
PSL, CNRS, Sorbonne Universit\'e, Universit\'e de Paris, 24 rue Lhomond, 75005 Paris, France}
\author{M. O. Goerbig}
\affiliation{Laboratoire de Physique des Solides, CNRS UMR 8502, Univ. Paris-Sud, Universit\'e
Paris-Saclay, F-91405 Orsay Cedex, France}
\author{B. Pla\c{c}ais} \email{bernard.placais@phys.ens.fr}
\affiliation{Laboratoire de Physique de l'Ecole normale sup\'erieure, ENS, Universit\'e
PSL, CNRS, Sorbonne Universit\'e, Universit\'e de Paris, 24 rue Lhomond, 75005 Paris, France}
\author{E. Baudin} \email{emmanuel.baudin@phys.ens.fr}
\affiliation{Laboratoire de Physique de l'Ecole normale sup\'erieure, ENS, Universit\'e
PSL, CNRS, Sorbonne Universit\'e, Universit\'e de Paris, 24 rue Lhomond, 75005 Paris, France}

\begin{abstract}

One of the intrinsic drift velocity limit of the quantum Hall effect is the collective magneto-exciton (ME) instability. It has been demonstrated in bilayer graphene (BLG) using noise measurements [W. Yang \textit{et al.}, Phys. Rev. Lett. \textbf{121}, 136804 (2018)]. We reproduce this experiment in monolayer graphene (MLG), and show that the same mechanism 
carries a direct relativistic signature on the breakdown velocity. Based on theoretical calculations of MLG- and BLG-ME spectra, 
we show that Doppler-induced instabilities manifest for a ME phase velocity determined by a universal value of the ME conductivity, 
set by the Hall conductance.

\end{abstract}

\maketitle

Low-bias quantum Hall (QH) transport is notoriously described in terms of single-electron physics, as exemplified by the edge-channel conductance quantization used in metrology [\onlinecite{Klitzing1980prl,Tzalenchuk2010nnano,Ribeiro2015nnano}]. The situation differs at large bias as electrons may couple to the collective particle-hole excitation spectrum (PHES) [\onlinecite{Goerbig2011rmp}], described by a dispersion relation $\omega(q)$ : it includes in the integer QH case both magneto-plasmon (MP) and magneto-exciton (ME) branches [\onlinecite{Roldan2009prb}], and, in the fractional QH case, a magneto-roton  (MR) 
branch [\onlinecite{Girvin1986prl,Jolicoeur2017prb}]. 
High-bias transport also differs in the electric field and current distributions. In a transistor or a Hall bar geometry (length $L$, width $W$), the non-dissipative Hall current penetrates the Landau insulating bulk, so that source and drain get connected via open ballistic orbits (drift velocity $v_x=E_y/B$) [\onlinecite{Streda1984jpc,Panos2014njp}]. The high-bias conductance $G_H$, and Hall conductivity  $\sigma_{xy}=G_H=I_x/V_y=\nu G_K$, are still set by the conductance quantum $G_K=e^2/h$ and the filling factor $\nu=nh/eB$ at a carrier density $n$. This ballistic transport is ultimately limited by the quantum Hall effect breakdown (QHEBD), a bulk effect  
occurring at a critical voltage $V_{bd}$ (or field  $E_{bd}=V_{bd}/W$, or velocity $v_{bd}=E_{bd}/B$), which is signaled by the onset of a longitudinal voltage  $V_x=LE_x$ associated with a bulk backscattering current and its associated shot noise $S_I$. The most frequently considered QHEBD mechanism is inter-Landau-level tunneling (ILLT), a single-particle effect that sets in when the wavefunctions of neighboring Landau-levels (LL) overlap in the tilted potential under applied bias.[\onlinecite{Eaves1986sst}]. In the case of a massive 2d electron gas called 2DEG,  ILLT has a critical Zener field $E_{Z}\sim\hbar\omega_c/eR_c$, where $\omega_c=eB/m^*$ and $R_c\sim\sqrt{N}l_B$ are the cyclotron angular frequency and radius, $m^*$ is the effective mass, $N$ the number of occupied LLs, and $l_B=\sqrt{\hbar/eB}$ the magnetic length [\onlinecite{Eaves1986sst}]. ILLT gives rise to quite large velocities $v_{Z}\sim \hbar/m^*R_c$ ($\sim2.10^5\;\mathrm{m.s^{-1}}$ for $N=1$ at $10~T$ with $m^*\simeq0.06~m_0$ for GaAs-based 2DEGs). Hall bar experiments indicate premature breakdowns with $v_{bd}\lesssim v_{Z}/10$ in both 2DEGs and graphene (see [\onlinecite{Yang2018prl}] and references therein). Several mechanisms have been considered to explain this discrepancy, such as the phonon- or impurity-assisted ILLT [\onlinecite{Eaves1986sst,Chaubet1998prb}]. Such extrinsic mechanisms are actually needed to overcome the momentum-conservation protection of ILLT, which stems from the $2k_F$ momentum mismatch between neighboring LL wave functions [\onlinecite{Dmitriev2012rmp}], where $k_F$ is the Fermi momentum. However, larger velocities $v_{bd}\sim v_Z/2$ have been reported in quantum-Hall constrictions [\onlinecite{Bliek1986sst,Chida2014prb}], thanks to a more uniform electrostatic landscape in the absence of invasive voltage probes. These experiments challenge the single-particle ILLT interpretation, and motivate alternative explanations in terms of collective excitations, such as the ME-instability scenario proposed in Ref. [\onlinecite{Yang2018prl}].

QHEBD was recently investigated in bilayer graphene (BLG) transistors using shot noise as a probe of ballistic transport breakdown [\onlinecite{Yang2018prl}]. Interestingly, doped BLG emulates a massive 2DEG with $m^*\simeq0.03~m_0$. In the  two-terminal transistor geometry, the breakdown was monitored by the sharp onset of the microwave shot-noise current $I_N=S_I/2e$ above the noiseless ballistic Hall background. Breakdown noise is characterized by a large differential noise conductance $G_N=\partial I_N/\partial V$ exceeding the DC Hall conductance $G_H$. These large values signal a strongly superpoissonian backscattering shot noise which has been interpreted in Ref.[\onlinecite{Yang2018prl}] as a signature of a collective magneto-exciton (ME) instability, calling for a kinematic origin of breakdown. The $\omega(q\sim k_F)$ sector of the 2DEG-PHES, which is relevant for breakdown in 2DEGs,  being essentially interaction independent (see [\onlinecite{Roldan2009prb}] and discussion below), the ME-instability velocity  $v_{ME}^{BLG}\sim\hbar/m^*R_c$ turns out to be similar to the interaction-free Zener limit $v_{Z}$, providing a clue to the apparent single-particle ILLT puzzle [\onlinecite{Yang2018prl}]. Even though the ME scenario can hardly be distinguished from ILLT  according to the breakdown threshold in 2DEGs, 
it does explain the superpoissonian noise as a mere consequence of its collective nature. Note that the ME-instability has also been considered to interpret quantum Hall fluid flows across an ionized impurity in Ref.[\onlinecite{Martin2003prl}], and DC magnetoresistance resonances in monolayer graphene (MLG) in Ref.[\onlinecite{Greenaway2021ncomm}].

The present work extends the noise investigation to MLG, that sustains a qualitatively different PHES
due to its relativistic Landau level ladder, and a more pronounced effect of interactions on the ME branches of the  PHES, 
as explained in Ref.[\onlinecite{Roldan2009prb}]. This peculiarity of MLG is revisited below, and in Supplementary Information Section III, with new RPA-calculations of the spectral function and magneto-optical conductivity $\sigma_{MO}$, accounting for screening by both hBN-encapsulation and local back-gating. Noise measurements, performed in high-mobility hBN-encapsulated graphene transistors, present a magnetic field and doping independent breakdown velocity $v_{bd}^{MLG}\simeq1.4\;10^5\;\mathrm{m/s}$. 
Calculations of the PHES for our transistor geometry indicate that this constant breakdown velocity is actually determined by an empirical but universal impedance matching criterion: $\sigma_{MO}\sim 10^{-2}~NG_K$, where $NG_K$ is the Hall conductance.  
We conclude the paper by a comparison between MLG and BLG breakdown velocities at large doping,
 illustrating this qualitative difference between massive and  massless ME-instability supported by RPA theory.

\begin{figure}[h!]
\centerline{\includegraphics[width=15cm]{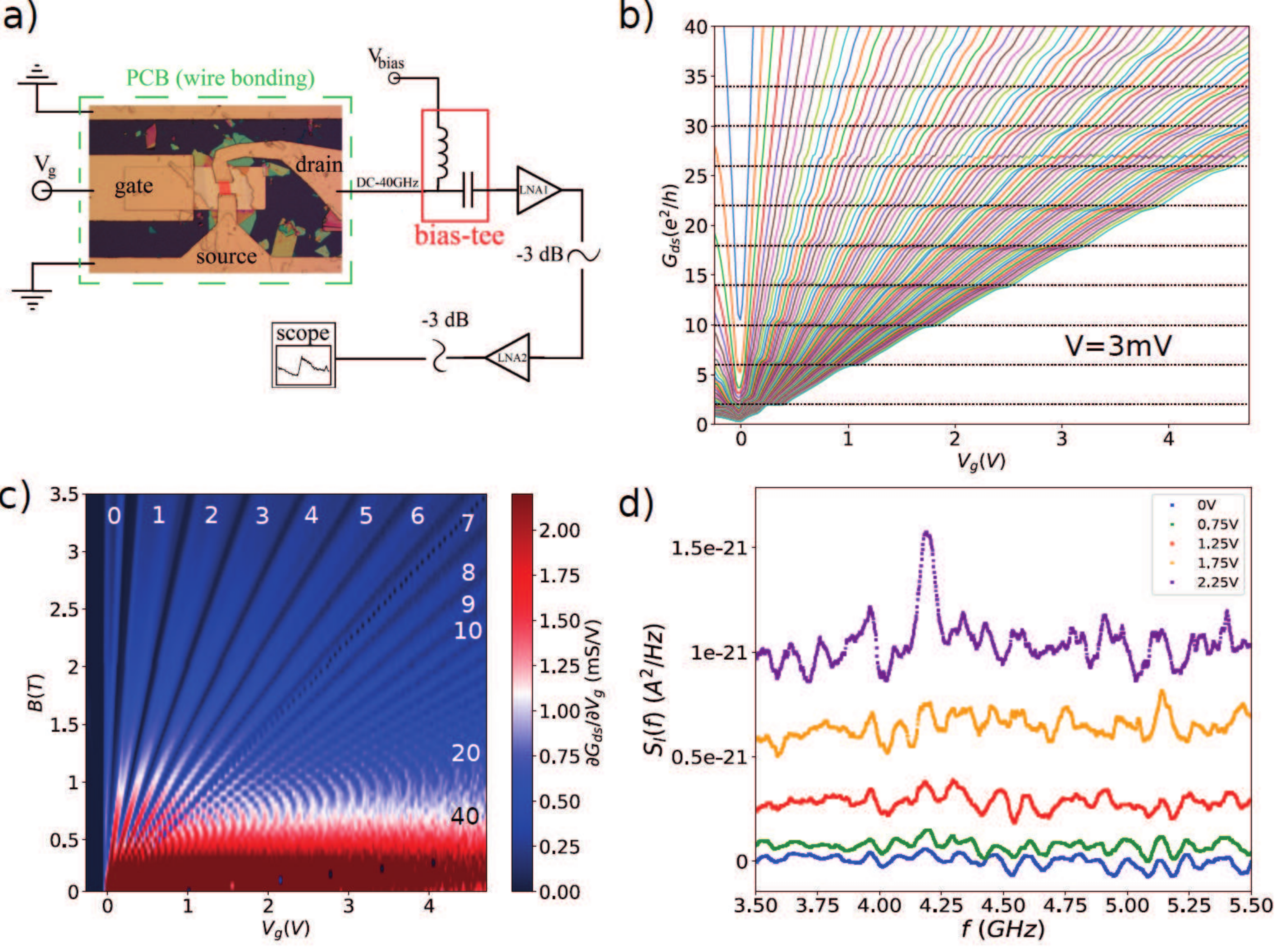}}
         \caption{Low-bias magneto-transport and noise in high-mobility bottom-gated hBN-encapsulated graphene sample AuS2 measured at $T=4\;\mathrm{K}$. Sample dimensions are $L\times W\times t_{hBN}=16\times10.6\times0.032\;\mathrm{\mu m}$. The contact resistance and mobility are $R_c=36\;\mathrm{Ohms}$ and $\mu=32\;\mathrm{m^2/Vs}$. a) sketch of the measuring setup. b) Low-bias ($V=3\;\mathrm{mV}$) conductance quantization steps in units of $e^2/h$, obeying the MLG quantization sequence $\nu=2(2N+1)$ for the filling factor $\nu$ as function of the Landau index $N$. c) Fan chart of the zero-bias differential conductance $\partial G_{ds}/\partial V_{g}$ showing a series of Landau levels. In b) and c), the feature inside the N=7 Landau level (at $G \simeq 27~e^2/h$) is a measurement artefact. d) Typical shot-noise spectra in increasing bias at $n=2.10^{12}\;\mathrm{cm^{-2}}$, $B=0.5\;\mathrm{T}$.  }
 \label{low-bias-AuS2.fig1}
\end{figure}

The samples analyzed in this experiment have been previously used in the investigations of the Schwinger effect in Ref.[\onlinecite{Schmitt2023nphys}] and/or flicker noise in Ref.[\onlinecite{Schmitt2023arXiv}]; they are described in Supplementary Information (Table.SI-1). The transistors are embedded in coplanar wave-guides for DC and microwave noise characterization at $4\;\mathrm{Kelvin}$ (see measurement setup in Fig.\ref{low-bias-AuS2.fig1}-a). The experiment is performed in the microwave frequency range to overcome flicker noise, which dominates up to the low-GHz range at large currents [\onlinecite{Schmitt2023arXiv}], and to access the QHEBD shot noise of interest. Data presented below concentrate on the hBN-encapsulated, bottom-gated, graphene sample AuS2 ($L \times W\times t_{hBN}=16\times10.6\times 0.032~\mu \mathrm{m}$) which is described in Fig.\ref{low-bias-AuS2.fig1} and in Ref.[\onlinecite{Schmitt2023nphys}]. Graphene conductance is calculated after correcting for the (small) contact resistance effect. Low-bias magneto-conductance $G(V_g,B)=\partial I/\partial V$ (Fig.\ref{low-bias-AuS2.fig1}-b), and the $\partial G/\partial V_g(V_g,B)$ fan-chart (Fig.\ref{low-bias-AuS2.fig1}-c), show clear MLG-quantization down to low fields, i.e. for $B\gtrsim 0.5\;\mathrm{T}$ in accordance with the large  $\mu\simeq32\;\mathrm{m^2/Vs}$ mobility. The specific MLG quantization, with plateaus at $\nu=2(2N+1)$, is clearly observed; the tiny width of the plateaus in Fig.\ref{low-bias-AuS2.fig1}-b signals the absence of disorder-induced localized bulk states, which warrants the absence of electrostatic-disorder. Plateaus gate voltages allow for the calibration of the gate capacitance at $C_g=1\;\mathrm{mF/m^2}$ for a thickness of the bottom-hBN $t_{hBN}=32\;\mathrm{nm}$ with $\epsilon_{hBN}=3.4$ [\onlinecite{Pierret2022MatRes}]. The large biases entail prominent drain-gating effects, eventually leading to a pinch-off, as reported in Ref.[\onlinecite{Schmitt2023nphys}] including for AuS2. This effect is compensated here by following the gating procedure described in Ref.[\onlinecite{Yang2018nnano}] and routinely used in Refs.[\onlinecite{Yang2018prl,Baudin2020adfm,Schmitt2023nphys,Schmitt2023arXiv}]; it consists in applying a bias-dependent gate voltage $V_{g}(V)=V_g(0)+ \beta V$, $\beta\sim0.4$ being adjusted to keep the resistance maximum at charge neutrality independent of bias at zero magnetic field. Fig.\ref{low-bias-AuS2.fig1}-d shows typical microwave $S_I(f)$ shot-noise spectra in increasing bias. Noise is expressed below in terms of the noise current $I_N=S_I/2e$ for an easy comparison with DC transport current.

\begin{figure}[h!]
\centerline{\includegraphics[width=17cm]{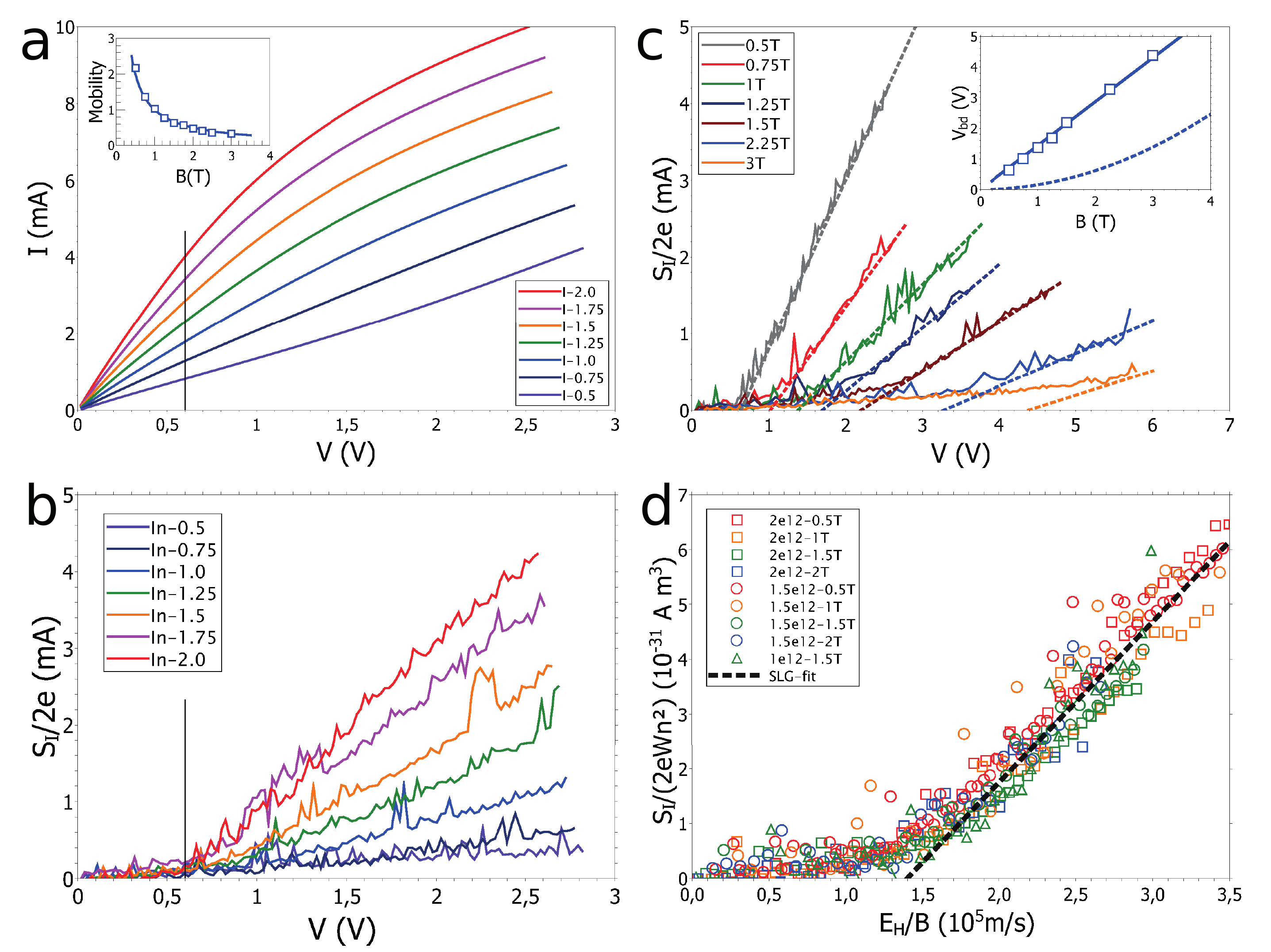}}
\caption{MLG magneto-transport and noise scaling in sample AuS2.
a) High-bias transport current $I(V)$ at $B=0.5\;\mathrm{T}$ for $n=0.5$-$2.10^{12}\;\mathrm{cm^{-2}}$ 
deviate from Quantum Hall current $I=ne V/B$ at the breakdown voltage $V_{bd}\simeq0.6\;\mathrm{V}$ (black line). Inset shows the measured low-bias mobility $I/neV$ (squares, in $\mathrm{m}^2/\mathrm{Vs}$) and the $1/B$ Hall line. 
b) Noise current $I_N(V)=S_I/2e$ at the same magnetic field and doping sequence, showing the onset of a large breakdown noise at the same
$V_{bd}$, above a low noise quantum Hall background $I_N/V\simeq 0.1\;\mathrm{mS}$ due to  contact noise. 
c) Voltage dependence of $I_N(V)=S_I/2e$ for various magnetic fields, at large doping $n=2.10^{12}\;\mathrm{cm^{-2}}$. Inset shows the linear dependence $V_{bd}(B)$ corresponding to $v_{bd}=0.14v_F$ (solid line), which strongly exceeds the Zener velocity (dashed line) at low field. 
d) Scaling of breakdown noise with Hall velocity $E_H/B$ for $n=1$-$2.10^{12}\;\mathrm{cm^{-2}}$ and $B=0.5$-$2\;\mathrm{T}$ 
corresponding to $V_{bd}=0.6$-$2.4\;\mathrm{V}$. It is represented by the master line $I_N/W=\gamma n^2[E_H/Bv_{bd}-1]$ (dashed line), with $v_{bd}(n,B)=1.4\;10^5\;\mathrm{m/s}$  (solid line), and $\gamma=4.10^{-31}\;\mathrm{A m^3}$ corresponding to large breakdown noise currents $I_N/W\sim40\;\mathrm{A/m}$ (at $n=1.10^{12}\;\mathrm{cm^{-2}}$). }
 \label{high-bias-Aus2.fig2}
\end{figure}

The high-bias magneto-transport and  noise characteristics of sample AuS2 are described in Fig.\ref{high-bias-Aus2.fig2}. The current voltage relation $I(V)$,
 measured at $B=0.5\;\mathrm{T}$ in Fig.\ref{high-bias-Aus2.fig2}-a, shows a smooth crossover between the  quantum Hall regime where
  $I\simeq I_H=\nu G_KV$ (inset) and the extremely high-bias metallic-like regime where the differential conductance recovers its zero-field value, which is set by the Zener-Klein conductivity [\onlinecite{Yang2018nnano,Yang2018prl}]. The breakdown voltage $V_{bd} \approx 0.6\;\mathrm{V}$ (black line) appears as a gradual deviation from the $I_H(V)$ Hall regime. By contrast, the current-noise characteristics $I_N(V)$, measured in the same conditions in Fig.\ref{high-bias-Aus2.fig2}-b, clearly distinguish two regimes : a quasi-noiseless quantum Hall regime for $V\leq V_{bd}=0.6\mathrm{V}$, characterized by a residual contact noise conductance $I_N/V\sim 0.1\;\mathrm{mS}$, and a large differential noise conductance $G_N(n)=\partial I_N/\partial V\gtrsim1\;\mathrm{mS}$ for $V\geq V_{bd}$. The intersection between the two lines provides an unambiguous determination of the breakdown voltage $V_{bd}$ which agrees with the transport determination in Fig.\ref{high-bias-Aus2.fig2}-a. In both $I(V)$ and $I_N(V)$, the breakdown voltage is found to be nearly doping-independent, 
as opposed to the high-bias noise conductance $G_N\propto n^2$ in Fig.\ref{high-bias-Aus2.fig2}-b. Fig.\ref{high-bias-Aus2.fig2}-c shows the current noise $I_N(V)$ for different magnetic fields at a fixed large doping $n=2.10^{12}\;\mathrm{cm^{-2}}$. It highlights the strong dependence of both $V_{bd}\propto B$ (inset) and $G_{N}\propto G_H\propto 1/B$, leading to a field-independent zero-bias extrapolate (not shown in the figure). These doping and field dependencies can be cast into the scaling displayed in Fig.\ref{high-bias-Aus2.fig2}-d, where noise data, collected over a broad $[n,B]$ range, are found to collapse on the universal master line
\begin{equation} \frac{I_N}{W}=\gamma n^2\left[\frac{E}{Bv_{bd}}-1\right]\qquad , \label{rouge} \end{equation} 
where $E=V/W$, $\gamma= 40 \times 10^{-32}~\mathrm{Am^3}$, and $v_{bd}= 1.4\times 10^5~\mathrm{m/s}$. 
While the scaling differs from that of BLG [\onlinecite{Yang2018prl}], the noise amplitudes are comparable, with $I_N/W=40~\mathrm{A/m}$ for $V=2~V_{bd}$ at $n=10^{12}\;\mathrm{cm^{-2}}$. This  noise scaling, with a doping-independent $v_{bd}$, contrasts with the doping-dependent ILLT breakdown threshold (dashed line in Fig. \ref{high-bias-Aus2.fig2}c-Inset). As quasi-particle interactions are controlled by the doping-independent fine-structure constant $\alpha_g=e^2/4\pi\hbar \epsilon_0\epsilon_r v_F$, the observation of a doping-independent $v_{bd}$ suggests a breakdown mechanism controlled by interactions. Besides, the current noise intensity $S_I\propto n^2$ corresponds to a doping-independent velocity noise $S_v\propto S_I/n^2$, suggesting 
a kinematic interpretation of breakdown such as that provided by the ME-instability. 

\begin{figure}[hh]
\centerline{\includegraphics[width=16cm]{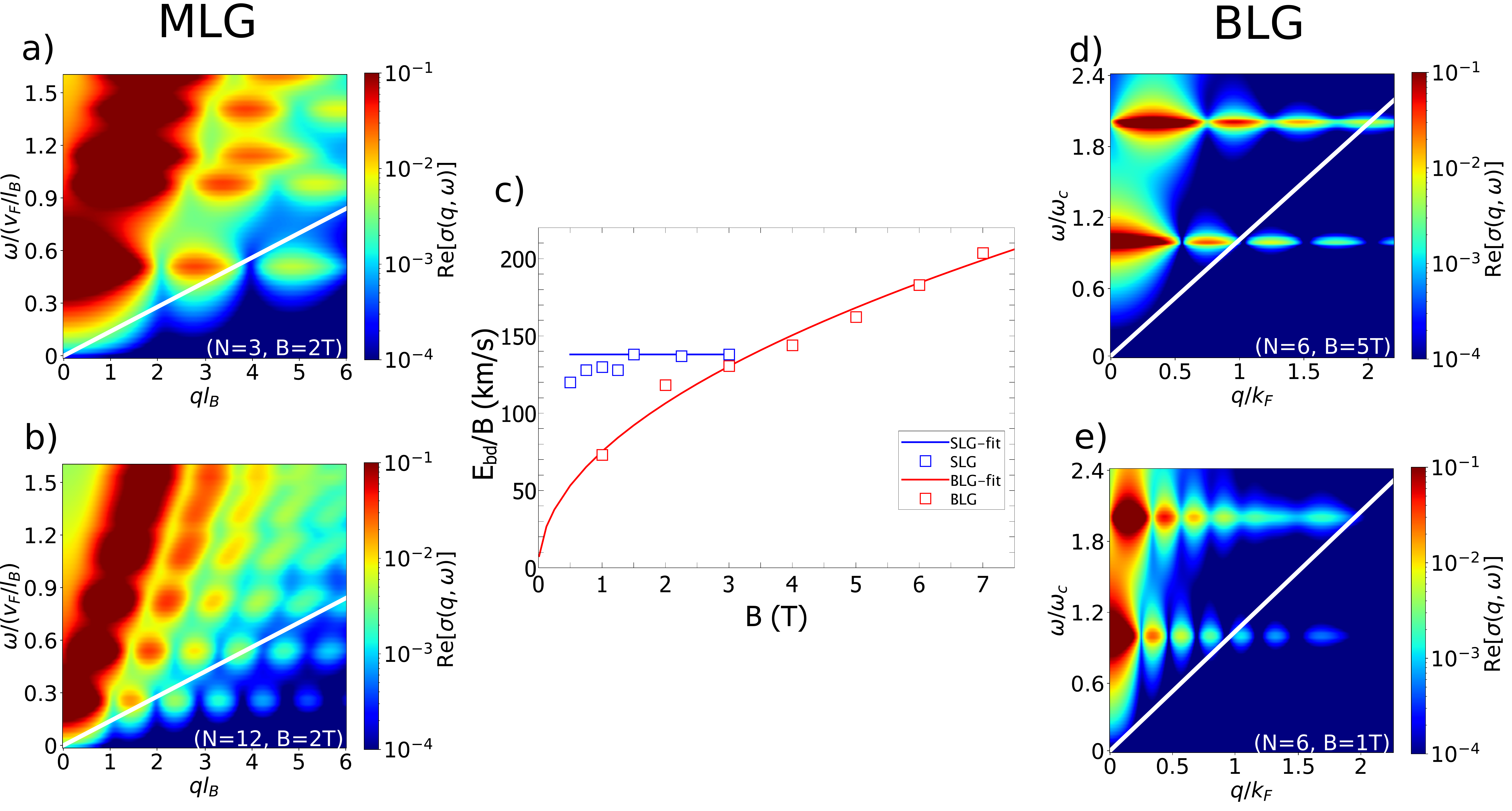}}
\caption{Effect of gate and dielectric screening on the velocity-induced magneto-exciton (ME) instability in MLG and BLG.
RPA magneto-optical conductivity spectra $\sigma_{MO}(q,\omega)$ (in units of $NG_K$) are plotted for the AuS2 MLG-sample geometry at
$B=2\;\mathrm{T}$ for $N=3$ in a) and $N=12$ in b), and a similar BLG sample at $N=6$ for $B=5\;\mathrm{T}$ in d) 
and $B=1\;\mathrm{T}$ in e). White lines correspond to the Doppler shifted drifting-electron spectrum for $v_{bd}^{MLG}=0.14 v_F$ in a) and b) 
and $v_{bd}^{BLG}=\hbar/m^*l_B\sqrt{N}\propto\sqrt{B}$ in d) and e). They separate the high-conductivity domain $\sigma_{MO}\gtrsim 10^{-2}\;NG_K$
from the low-conductivity one. 
c) Experimental data of $v_{bd}^{MLG}$ (blue squares) and $v_{bd}^{BLG}$ (red squares from Ref.[\onlinecite{Yang2018prl}]),
measured at the same $n=2.10^{12}\;\mathrm{cm^{-2}}$, support these scalings and illustrate the main difference between MLG and BLG.}
\label{phes.fig3}
\end{figure}

To base this qualitative interpretation on a more quantitative analysis, we recalculate below the MLG-PHES of Ref.[\onlinecite{Roldan2009prb}], 
adapting it for geometry and material parameters that are suitable for our experimental conditions. 
Figures \ref{phes.fig3}-(a,b) show the PHESs calculated in the RPA approximation of Ref.[\onlinecite{Roldan2009prb}]. It is adapted for AuS2-sample geometry by including the screening by the local bottom-gate and the hBN-encapsulation, as explained in Supplementary Section-III.A. In the context of velocity-induced instability, we have plotted the magneto-optical conductivity spectrum $\Re [\sigma_{MO}(q,\omega)]$ (denoted $\sigma_{MO}$ below), which is deduced from the usual spectral function $\Im [\Pi^{RPA}(q,\omega)]$ using the $\Re [\sigma_{MO}(q,\omega)] = -\frac{\omega e^2}{q^2} \Im [\Pi^{RPA}(q,\omega)]$ relation. Note that  $\sigma_{MO}$ suffers from non-physical divergence in the low-$q$ PHES-limit, blurring the magneto-plasmon branch and hiding the existence of a spectral gap that appears more clearly in the spectral function (see $\Pi^{RPA}$ spectra in Supplementary Fig. SI-5). This $N$-dependent bandgap is equal to the MLG cyclotron gap $\omega_c^{MLG}=(\sqrt{N+1}-\sqrt{N})v_F/l_B \simeq v_F/R_c$ (with $R_c$ the cyclotron radius), similarly to BLG in Figs.\ref{phes.fig3}-(d,e), where $\omega_c^{BLG}=1/m^*l_B^2$ is $N$-independent (see the spectral function in Fig. SI-7). Conductivity spectra are plotted for $N=3$ (panel a) and $N=12$ (panel b), displayed 
in logarithmic scale to map their steep $\omega$ and $q$ dependencies, and normalized to the Hall conductivity $NG_K$ (per spin and valley) for a direct comparison of electronic and collective electron-hole excitation's conductivity. The momentum and energy scales are expressed in MLG-relevant dimensionless units $ql_B$ and $\omega l_B/v_F$, which imply a magnetic-field independence of the ME branches phase velocity $v_{ME}=\omega_{ME}/q$. Remarkably the ME optical conductivity $\sigma_{MO}$ is  steeply increasing with $v_{ME}$, with $\sigma_{MO}\sim(3.10^{-4}$-$3.10^{-2})\;NG_K$ for $v_{ME}=(0.06$-$0.35)\;v_F$.  

The effect of screening is quite substantial in MLG, as depicted in Supplementary Section Fig.SI-6, 
and much more prominent than in BLG (Fig.SI-8), especially at large $q$. The white lines in Figs.\ref{phes.fig3}-(a,b) correspond to the Doppler shifted electronic energy $\omega=v_{bd}q$ of drifting electrons, calculated at the  measured breakdown velocity $v_{bd}=0.14v_F$ of Fig.\ref{high-bias-Aus2.fig2}. In both the $N=3$ (panel a) and the $N=12$ (panel b) examples, this line separates a high ME-conductivity domain for $\omega\gtrsim v_{bd}q$, where $\sigma_{MO}\gtrsim10^{-2}NG_K$, from a low conductivity domain for $\omega\lesssim v_{bd}q$. The observation of a $N$- and $B$-independent ME-instability, at a velocity $v_{BD}\simeq v_{ME}=\mathrm{cst.}$ controlled by a fixed $\sigma_{MO}\sim10^{-2}NG_K$ constraint, is consistent with a collective wave interpretation, 
even if the  value of the impedance threshold remains to be established theoretically. It is obviously consistent with our experimental observation in Fig.\ref{high-bias-Aus2.fig2} of an $n$- and $B$-independent breakdown velocity, 
a feature observed in all tested Au-gated samples (see Supplementary Table SI-1). 
Unlike in BLG, the $v_{bd}=\mathrm{cst.}$ breakdown velocity of MLG, inferred from the above conductivity criterion, 
exceeds the  ILLT  $v_Z\propto\sqrt{B}$, especially at low-$B$. 

Let us recall that the situation is different in BLG. Figures \ref{phes.fig3}-(d,e) reproduce the theoretical analysis for a similar BLG sample, 
such as that measured in Ref.[\onlinecite{Yang2018prl}]. The energy ($\omega/\omega_c$) and momentum ($q/k_F$) reduced units are adapted for a massive 2DEG-like BLG, but the reduced conductivity scale $\sigma_{MO}/NG_K$ is the same. The two panels correspond to the same $N=6$, but different magnetic fields $B=5\;\mathrm{T}$ (panel d) and $B=1\;\mathrm{T}$ (panel e). Contrarily to MLG, the phase velocity $\omega_{ME}/q$ is not magnetic-field independent in this representation, as $\omega_c \propto B$ and $k_{F} \propto 1/l_B \propto \sqrt{B}$ have different B-dependencies.  As a consequence, positioning an identical Doppler line on the two reduced-units plots 
amounts taking a $v_{bd}\propto\sqrt{B}$. Figs.\ref{phes.fig3}-(d,e) show that this criterion also corresponds to 
a consistent $\sigma_{MO}\sim10^{-2}NG_K$ criterion for the ME-instability, which is met in BLG at ($q,\omega$)-localized ME conductivity peaks. This impedance analysis shows that for BLG, and more generally 2DEGs, the ME-instability and Zener-ILLT, which are basically different, give consistent and similar breakdown velocities in BLG, 
confirming earlier statement of Ref.[\onlinecite{Yang2018prl}]. Finally, Fig.\ref{phes.fig3}-c illustrates the qualitative difference between MLG and BLG in a plot of $v_{bd}(B)$ at a large $n=2.10^{12}\;\mathrm{cm^{-2}}$ (BLG data are reproduced from Fig.2 of Ref.[\onlinecite{Yang2018prl}]), with  $v_{ME}^{MLG}\simeq 0.14 v_F$ (blue line) and $v_{ME}^{BLG}=\hbar/m^*l_B\sqrt{N}\propto\sqrt{B}$ (red line for $N=5$) [\onlinecite{Yang2018prl}]. Let us note that the magnetic field dependencies $v_{bd}^{MLG}=\mathrm{Cst.}$ and $v_{bd}^{BLG} \propto \sqrt{B}$ merely reflect the energy dependence of the Fermi velocity, $v_F^{MLG}(\varepsilon_L)=\mathrm{Cst.}$ and $v_F^{BLG}(\varepsilon_L) \propto \sqrt{\varepsilon_L}$, when taken at the Landau energy $\varepsilon_L = \hbar \omega_c \propto B$.

Relying on the good mapping of the ME-scenario with experiment, we exploit the RPA calculations further in Supplementary Section III-B and III-C to model breakdown in varied graphene geometries such as graphene 
in vacuum or semi-infinite hBN embedding, keeping a systematic benchmark of MLG and BLG cases, and assuming the existence of a universal impedance-matching condition. 
For MLG, we show in Figs.SI-6-(a,b) that screening by the bottom gate in AuS2 (panel a) is equivalent to a semi-infinite, 
$\epsilon_r=100$ dielectric (panel b), meaning that both PHESs correspond to the fully screened conductivity. 
Effect of interactions, which is maximal for un-gated suspended graphene ($\epsilon_r=1$ in panel c), amounts to suppressing the conductivity amplitude below the 
$\sigma_{MO}\sim10^{-2}NG_K$ ME-instability threshold over most of the 
ME-spectrum leading to an enhanced breakdown velocity $v_{bd}\simeq 0.5 v_F$ (white line). Given the impedance matching 
condition $\sigma_{MO}\sim10^{-2}NG_K$, we conclude that the ME-instability velocity of MLG is a constant, 
in the range $v_{ME}=[0.14,0.5]\; v_F$ that depends on screening.    
The same analysis is performed for BLG in Figs.SI-8, showing that the  large-$q$ PHES sector is to a large extent insensitive to screening, yielding a Zener-like breakdown velocity  $v_{ME}^{BLG}=\frac{\hbar}{m^*R_c}$. 

In conclusion, we have shown that bulk quantum Hall breakdown is controlled by the magneto-exciton instability in both MLG and BLG with a threshold $v_{drift} \geq v_{ME}$, which is reminiscent of the Cerenkov effect [\onlinecite{Landau1969MIR}]. More precisely, instability is defined by a universal conductivity criterion $\sigma_{MO}\sim10^{-2}NG_K$. This universal criterion explains the qualitative differences between the massless MLG, and massive BLG. Whereas the BLG-ME instability mimics single-particle ILLT, 
that of MLG is sensitive to screening by the embedding dielectric and local gates. Screening reduces the breakdown velocity, and gated transistors correspond to the fully screened regime. Both studies promote shot-noise as a sensitive probe of quantum Hall transport, 
RPA as a relevant theoretical tool to tackle interactions and screening, 
and high-velocity transport as a sensitive  probe of the large-momentum collective excitations, as suggested by Landau [\onlinecite{Landau1941pr}]. 
Understanding the combined effects of Landau quantization and interactions in the collective modes of the integer quantum Hall effect is a prerequisite before 
addressing the more challenging case of the fractional regime, where elusive magneto-rotons may come into play. Finally and on a broader scope, let us mention that the magneto-exciton instability is a  quantum-Hall-matter light coupling effect, which belongs to a domain of current interest [\onlinecite{Appugliese2022science}].

\section*{Supplementary Information}
A Supplementary Information is available. It presents a similar analysis of the experimental data on the other devices of the series, as well as an extended discussion on the magneto-optical conductivity spectrum in monolayer and bilayer graphene. In this respect, it focuses on the contrasted role of interactions in these two cases.

\begin{acknowledgments} AS thanks Prof. C.R. Dean for hospitality and introducing him to the fabrication of high-quality graphene-hBN heterostructures.  The research leading to these results has received partial funding from the European Union Horizon 2020 research and innovation program under grant agreement No.881603 "Graphene Core 3", and from the French  ANR-21-CE24-0025-01 "ELuSeM". 
\end{acknowledgments}

\section*{Conflict of interest}
The authors have no conflict of interest to disclose.

\section*{Authors contribution statement}
AS, BP and EB conceived the experiment. AS conducted device fabrication and measurements, under the guidance of MR in the early developments. TT and KW have provided the hBN crystals. AS, MOG and BP developed the models and theoretical interpretations. AS, GF, JMB, GM, CV, BP and EB participated to the data analysis. BP wrote the manuscript with assistance of AS and EB, and contributions from the coauthors. 

\section*{Data availability statement}

Data are available on a public Zenodo repository.

\newpage

\end{document}


\title{Magneto-exciton limit of quantum Hall breakdown in graphene :  Supplementary Information}

\author{A. Schmitt}\email{aurelien.schmitt@phys.ens.fr}
\affiliation{Laboratoire de Physique de l'Ecole normale sup\'erieure, ENS, Universit\'e
PSL, CNRS, Sorbonne Universit\'e, Universit\'e de Paris, 24 rue Lhomond, 75005 Paris, France}
\author{M. Rosticher}
\affiliation{Laboratoire de Physique de l'Ecole normale sup\'erieure, ENS, Universit\'e
PSL, CNRS, Sorbonne Universit\'e, Universit\'e de Paris, 24 rue Lhomond, 75005 Paris, France}
\author{T. Taniguchi}
\affiliation{Advanced Materials Laboratory, National Institute for Materials Science, Tsukuba,
Ibaraki 305-0047,  Japan}
\author{K. Watanabe}
\affiliation{Advanced Materials Laboratory, National Institute for Materials Science, Tsukuba,
Ibaraki 305-0047, Japan}
\author{G. F\`eve}
\affiliation{Laboratoire de Physique de l'Ecole normale sup\'erieure, ENS, Universit\'e
PSL, CNRS, Sorbonne Universit\'e, Universit\'e de Paris, 24 rue Lhomond, 75005 Paris, France}
\author{J.M. Berroir}
\affiliation{Laboratoire de Physique de l'Ecole normale sup\'erieure, ENS, Universit\'e
PSL, CNRS, Sorbonne Universit\'e, Universit\'e de Paris, 24 rue Lhomond, 75005 Paris, France}
\author{G. M\'enard}
\affiliation{Laboratoire de Physique de l'Ecole normale sup\'erieure, ENS, Universit\'e
PSL, CNRS, Sorbonne Universit\'e, Universit\'e de Paris, 24 rue Lhomond, 75005 Paris, France}
\author{C. Voisin}
\affiliation{Laboratoire de Physique de l'Ecole normale sup\'erieure, ENS, Universit\'e
PSL, CNRS, Sorbonne Universit\'e, Universit\'e de Paris, 24 rue Lhomond, 75005 Paris, France}
\author{M. O. Goerbig}
\affiliation{Laboratoire de Physique des Solides, CNRS UMR 8502, Univ. Paris-Sud, Universit\'e
Paris-Saclay, F-91405 Orsay Cedex, France}
\author{B. Pla\c{c}ais} \email{bernard.placais@phys.ens.fr}
\affiliation{Laboratoire de Physique de l'Ecole normale sup\'erieure, ENS, Universit\'e
PSL, CNRS, Sorbonne Universit\'e, Universit\'e de Paris, 24 rue Lhomond, 75005 Paris, France}
\author{E. Baudin} \email{emmanuel.baudin@phys.ens.fr}
\affiliation{Laboratoire de Physique de l'Ecole normale sup\'erieure, ENS, Universit\'e
PSL, CNRS, Sorbonne Universit\'e, Universit\'e de Paris, 24 rue Lhomond, 75005 Paris, France}

\begin{abstract}

This Supplementary Information has two goals. On one hand, it gives a broader description of the other devices of the series and establishes the reproducibility of the main text analysis performed on AuS2 device. On the other hand, it gives a detailed description of the RPA calculations of the magneto-optical conductivity for monolayer and bilayer graphene, and puts the emphasis on the contrasted role played by the interactions.

\end{abstract}

\maketitle

\renewcommand{\thefigure}{SI-\arabic{figure}}  
\renewcommand{\thetable}{SI-\arabic{table}}  

\section{Description of the samples}
Apart from the sample AuS2 extensively described in the main text, 3 other high-mobility hBN-encapsulated graphene transistors have been characterized. Table \ref{tab:label} presents their relevant properties. The 4 transistors present a large size $L,W > 6 \mu\mathrm{m}$, an important mobility $\mu(0) \gtrsim 15 \mathrm{m}^{2}/\mathrm{Vs}$ and contrasted thickness of hBN dielectric $t_{hBN} =[34, 152] \mathrm{nm}$. All of them present transparent edge contacts, well suited for GHz noise characterization. Using the same scaling as the one presented in Figure 2-d of the main text, all of them exhibit a doping- and magnetic field-independent breakdown velocity, whose values are presented in the last column of Table \ref{tab:label}. 

\begin{table}[h]
\centering
\hspace*{-0.7cm}
    \begin{tabular}{| p{2.0cm} |  c | c | c | c | c | c |  c |  c |  c |  c |  c |c|c|}
    \hline
    \textbf{Sample} & \textbf{Gate} & \textbf{L} & \textbf{W}& \boldmath{$t_{hBN}$} & \boldmath{$R_{c}$} & \boldmath{$\;\mu(0)\;$}&\boldmath{$\;v_{bd}\;$} \\
       name & & $\;\mathrm{\mu m}\;$ &
    $\;\mathrm{\mu m}\;$ & $\mathrm{nm}$ & $\;\;\;\Omega\;\;\;$ & $\;\mathrm{m^{2}/Vs }   $ &$\;10^6\;\mathrm{m/s} $ \\ \hline\hline
    \textbf{AuS2}& Au & 16 & 10.6 & 34 & 36 & \textbf{32} & 0.14  \\ \hline
    \textbf{AuS3}& Au & 11.1 & 11.4 & 90 & 75 & \textbf{14}  & 0.16 \\ \hline
    \textbf{Flicker161}& Au & 6 & 10 & 152 & 21  & \textbf{28} & 0.09 \\ \hline
    \textbf{Heinrich}& Graphite & 15 & 10 & 53 & 69  & \textbf{38} & 0.05  \\ \hline
    \end{tabular}
\caption{Geometrical and electrical properties of the $4$ devices series. Device AuS2 is analyzed in the main text.  Basic properties include the nature of the gate electrode (Au or Graphite bottom gating), the channel length $L$, width $W$,  hBN dielectric thickness $t_{hBN}$, and contact resistance $R_{c}$. The mobility $\mu(0)$ is extracted from transfer curves measured at $T=4K$. The last column contains the value of the magnetic field- and doping-independent breakdown velocity extracted from the scaling presented in Figure 2-d of the main text.}
\label{tab:label}
\end{table}

\section{Quantum Hall breakdown in the complete series}

\begin{figure}[h!]
\centerline{\includegraphics[width=16cm]{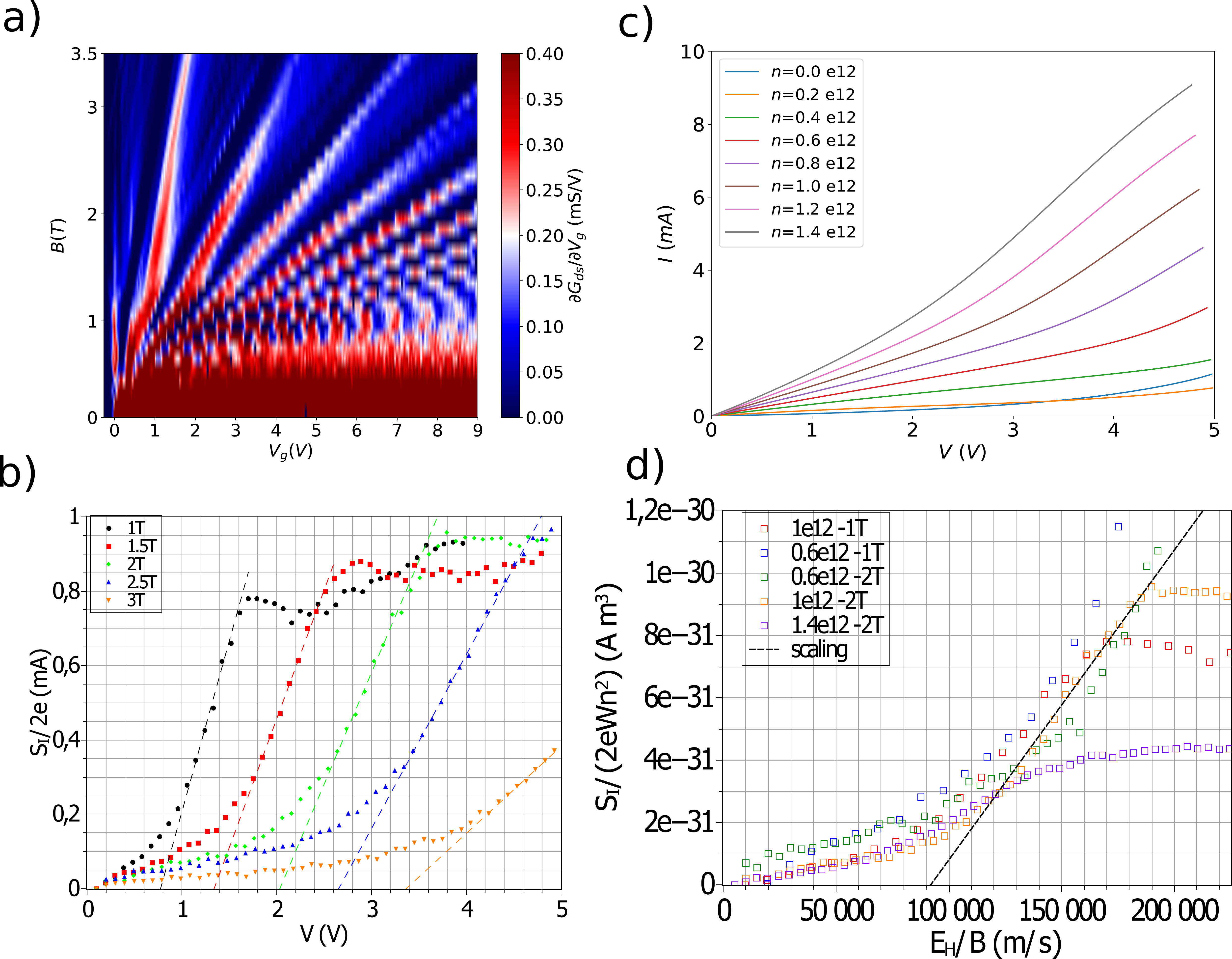}}
\caption{Two-terminal magneto-transport and noise in the high-mobility, hBN-encapsulated, bottom-gated sample Flicker161. a) Fan chart of the zero-field differential conductance $\partial G_{ds}/\partial V_{g}$ showing a series of Landau levels and the lifting of their 4-fold degeneracy for the $N=0$ state. c) High-bias transport current $I(V)$ at $B=2\;\mathrm{T}$ for different doping $n=0$-$1.4~10^{12}\;\mathrm{cm^{-2}}$. Quantum Hall breakdown corresponds to a weak deviation from the ballistic Hall transport $I=ne V/B$. b) Bias dependence of shot-noise at large doping $n=1.10^{12}\;\mathrm{cm^{-2}}$ in increasing magnetic field $B=1$-$3\;\mathrm{T}$. The breakdown voltage, $V_{bd}(B)=0.8$-$3.4\;\mathrm{V}$, strongly depends on magnetic field. d) Scaling of the breakdown noise with Hall velocity $E_H/B$ for $n=0.6$-$1.4~10^{12}\;\mathrm{cm^{-2}}$ and $B=1$-$2\;\mathrm{T}$. The scaling obeys the master line $I_N/W=\beta n^2[E_H/Bv_{bd}-1]$ (dashed line), with $v_{bd}(n,B)=0.9\;10^5\;\mathrm{m/s}$ }
 \label{F161}
\end{figure}

In this section we provide additional analysis on the other devices of the series. Let us first concentrate on the device Flicker161, whose properties are summarized in Table \ref{tab:label} and which is described in Figure \ref{F161}. 

This device presents well-defined Landau levels, with very narrow conductance plateaus (see panel a), highlighting a very clean sample. Its high-bias transport current $I(V)$, presented in panel b, shows a behavior qualitatively similar to the one observed in AuS2 device, with a deviation from the Hall transport at high bias signaling the QHE breakdown. As underlined in the main text, QHE breakdown is more easily visible when looking at the bias dependence of shot noise: such behavior is plotted in panel c) for a fixed value of doping and a variety of magnetic field values, with a breakdown voltage, signaled by the steep increase of the noise, that strongly depends on magnetic field. The observed breakdown noise obeys the same scaling as the one reported in Figure 2-d of the main text, demonstrating a doping- and magnetic field-independent breakdown velocity $v_{bd} = 0.9~10^{5}\mathrm{m/s}$ in this device.

  \begin{figure}[h!]
\centerline{\includegraphics[width=16cm]{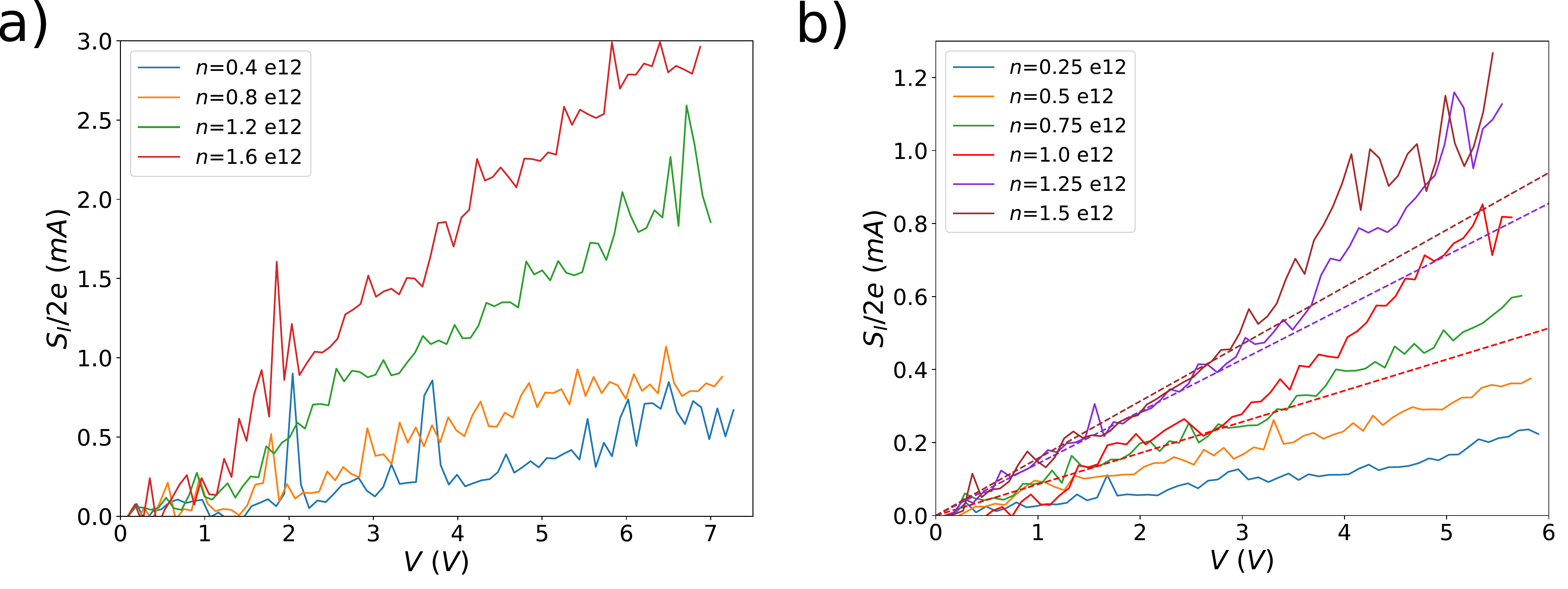}}
\caption{Bias dependence of the measured shot noise on the two other devices of the series, for a variety of doping. Panel a) shows the Heinrich device at $B=1.5\mathrm{T}$, panel b) corresponds to the AuS3 device at $B=1.5\mathrm{T}$. For the latter, the change of slope associated to the breakdown is less pronounced, but is however visible (see the guidelines for high doping values).}
 \label{autres}
\end{figure}

\begin{figure}[h!]
\centerline{\includegraphics[width=12cm]{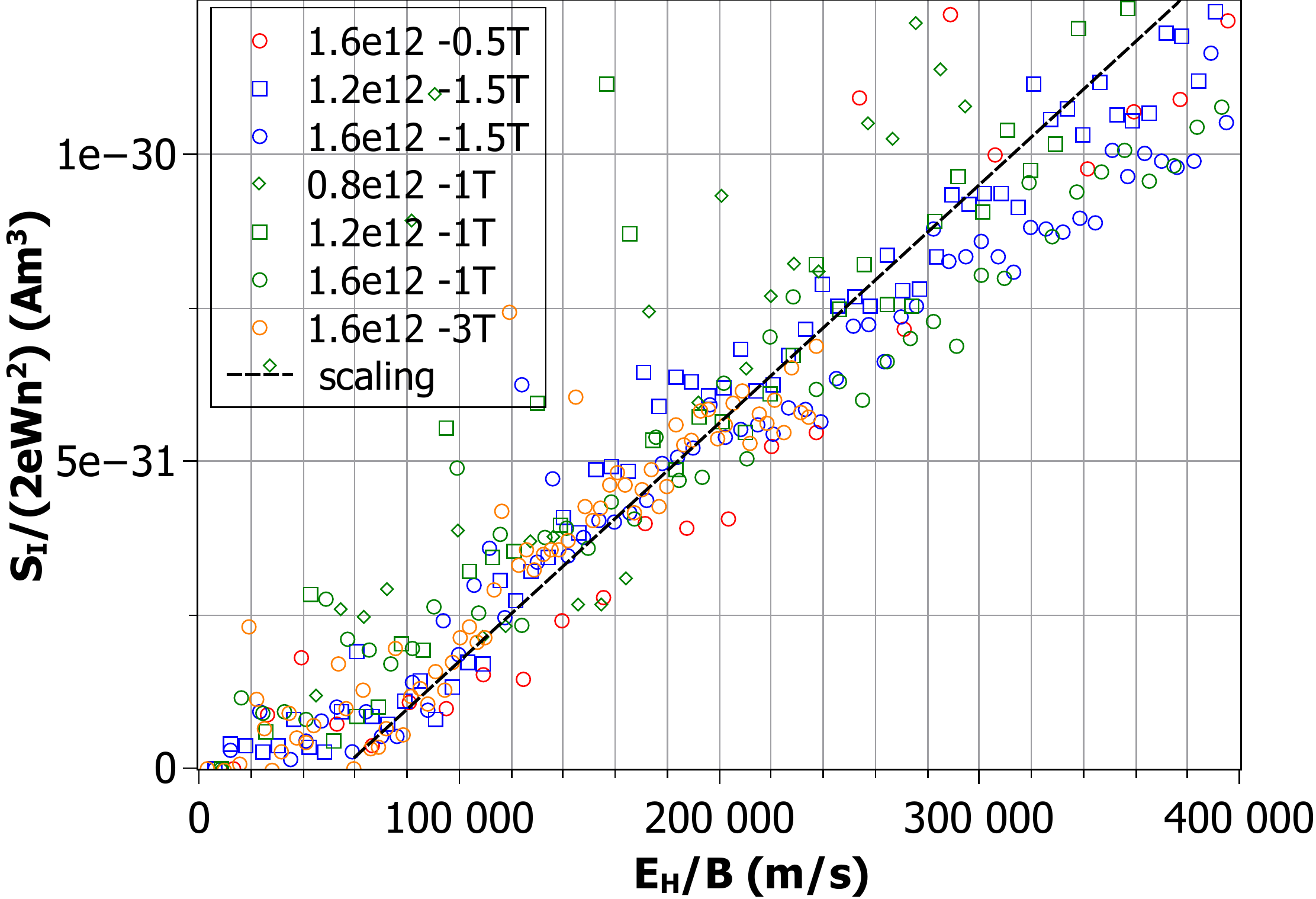}}
\caption{Breakdown noise scaling for the Heinrich device, for a variety of doping $n=[0.8,1.6]~10^{12}~\mathrm{cm}^{-2}$ and a magnetic field range $B=[0.5,3]~\mathrm{T}$}
 \label{Heinrich}
\end{figure}

The two other devices have been analyzed in the same way: a glimpse on the bias dependence of their shot noise is presented in Figure \ref{autres}. Once again, the QHE breakdown is signaled by a strong increase of the shot-noise, and a complete analysis reveals a doping- and magnetic field-independent breakdown velocity, whose values are summarized in Table \ref{tab:label}. As an example, we represent in Figure \ref{Heinrich} a similar scaling of the breakdown noise for the Heinrich device, for a large variety of dopings and fields.

\section{Particle-hole excitation spectrum in the RPA approximation}

\subsection{General framework}

The determination of the particle-hole excitation spectrum in the case of monolayer and bilayer graphene is realized following the method developed in Ref.[\onlinecite{Roldan2009prb}]: for the sake of clarity, we partially reproduce their derivation below. In the case of bilayer graphene, we rely on the calculations for a conventional 2DEG, that match the bilayer case in the large N limit [\onlinecite{Goerbig2009}], where N is the last occupied Landau level.

In Ref.[\onlinecite{Roldan2009prb}], RPA approximation is applied by considering an unscreened 2D Coulomb potential that reads: 
\begin{equation}
    v(q)=\frac{2\pi e^2}{\epsilon_b q}
\label{potFAKE}
\end{equation}
where $\epsilon_b$ is the dielectric permittivity of graphene environment.

\begin{figure}[h!]
\centerline{\includegraphics[width=9cm]{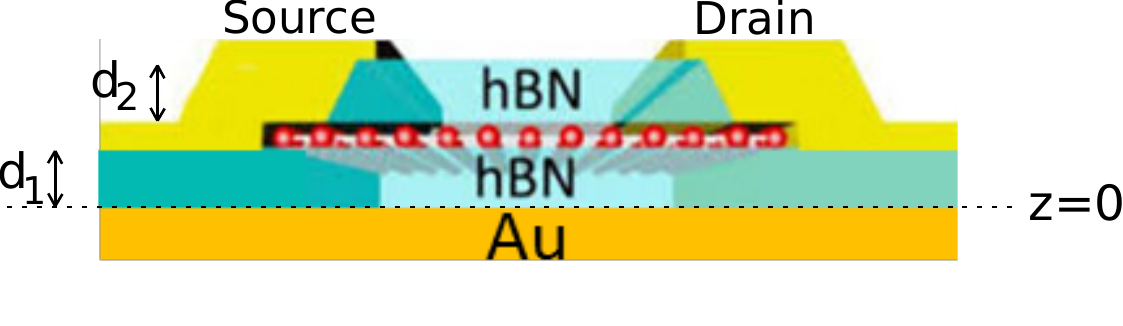}}
\caption{Heterostructure consisting of a hBN/graphene/hBN stack on top of an Au backgate.}
 \label{dessin_dispo}
\end{figure}

In a more realistic approach, we consider for the RPA approximation a screened 2D Coulomb potential $v(q,\omega)$ accounting for the presence of a local bottom-gate and for the hBN encapsulation. In this respect, we compute the electrostatic interaction within a vertical heterostructure composed of: 1) a graphene sheet ; 2) a top hBN layer of thickness $d_{2}$ ; 3) a bottom hBN layer of thickness $d_{1}$ (which corresponds to the hBN thickness denoted $t_{hBN}$ in Table \ref{tab:label}). The whole structure is on top of a Au backgate located at $z=0$, and the half-space $z>(d_{1}+d_{2})$ is filled with air. The hBN slabs are associated to frequency-dependent dielectric functions $\epsilon_{x}(\omega)$ and $\epsilon_{z}(\omega)$ for in-plane and out-of-plane components respectively. A schematics is presented in Figure \ref{dessin_dispo}.

In a similar manner to Ref. [\onlinecite{Principi2017prl}], and in accordance with the unscreened limit when $d_1 \rightarrow \infty$, we obtain the following expression for the potential : 

\begin{equation}
    v(q,\omega)=\frac{4 \pi e^2 ~\mathrm{sinh}\left[d_{1}\sqrt{\frac{\epsilon_x}{\epsilon_z}}q\right] \bigg( \sqrt{\frac{\epsilon_x}{\epsilon_z}} q \epsilon_z ~\mathrm{cosh}\left[d_{2} \sqrt{\frac{\epsilon_x}{\epsilon_z}} q \right] +q ~\mathrm{sinh}\left[d_{2} \sqrt{\frac{\epsilon_x}{\epsilon_z}} q \right] \bigg)  }{\sqrt{\frac{\epsilon_x}{\epsilon_z}} q \epsilon_z \bigg( \sqrt{\frac{\epsilon_x}{\epsilon_z}} q \epsilon_z ~\mathrm{cosh}\left[(d_1+d_2)\sqrt{\frac{\epsilon_x}{\epsilon_z}}q \right] +q ~\mathrm{sinh}\left[(d_1+d_2)\sqrt{\frac{\epsilon_x}{\epsilon_z}}q\right] \bigg)} 
\label{potREEL}
\end{equation}
where the $\omega$-dependence appears through the frequency dependence of the dielectric properties $\epsilon_x$ and $\epsilon_z$ of the hBN encapsulant. A low-$q$ development of this full potential expression yields : 
\begin{equation}
    v(q,\omega) \simeq \frac{4 \pi e^2 d_1 }{\epsilon_z} \biggl(1- \frac{d_{1}q \epsilon_0}{\epsilon_z} \biggr)
\end{equation}
which coincides to leading order with the potential $v_\text{scr}(q)=(2\pi e^2/\epsilon_z q)[1-\exp(-2d_1q)]\sim 4\pi e^2 d_1/\epsilon_z$ of an electric point charge in the presence of a metallic gate at a distance $d_1$ from the graphene sheet, embedded in a dielectric environment described by the dielectric constant $\epsilon_z$.

Expressions for the bare polarization function $\Pi^{0}(q,\omega)$ for MLG and BLG are detailed in the dedicated sections below. From this non-interacting polarization, we get the renormalized polarization function within the RPA approximation : 

\begin{equation}
    \Pi^{RPA}(q,\omega)=\frac{\Pi^{0}(q,\omega)}{1-v(q,\omega) \Pi^{0}(q,\omega)}
\label{EqRPA}
\end{equation}
The latter can be converted into the magneto-optical conductivity  through the relation [\onlinecite{Bruus}] : 
\begin{equation}
\Re [\sigma_{MO}(q,\omega)] = -\frac{\omega e^2}{q^2} \Im [\Pi^{RPA}(q,\omega)] \qquad ,
\end{equation}
which is plotted in Figure 3 of the main text,

\subsection{Monolayer graphene}

Let us first recall the main ingredients for the derivation of the polarizability of monolayer graphene in a magnetic field, as described in Ref. [\onlinecite{Roldan2009prb}]. Monolayer graphene is associated to a relativistic Landau level spectrum $E_n$ : $E_n=\lambda \epsilon_n=\lambda \frac{v_F}{l_B} \sqrt{2n}$ where $l_B=\sqrt{\hbar/eB}$ is the magnetic length and $\lambda$ denotes the band index ($\lambda =\pm 1$ for conduction/valence band); associated eigenstates are spinors. In order to calculate the polarization function for monolayer graphene, it is thus necessary to compute the single-particle Green's function, which is a $2\times2$ matrix due to the spinor character of the wavefunction. 

We first calculate the matrix element associated to the density operator : 
\begin{equation}
    \langle n | e^{i\mathbf{q}  \mathbf{r}}| n_{1} \rangle = e^{-l_B^2 \mathbf{q}^2/4} \biggl[\Theta (n-n_1) G(n,n_1,q_x,q_y,B) + \Theta(n_1-n-1) G(n_1,n,q_x,-q_y,B)\biggr]
\end{equation}
with $\Theta$ the unit step function, $| n \rangle$ the eigenstates for MLG in a magnetic field, and the function G is defined by : 

\begin{equation}
    G(n,n_1,q_x,q_y,B)=\biggl(\frac{i l_B}{\sqrt{2}}(q_x-iq_y) \biggr)^{n-n_1} \sqrt{\frac{(n_1 ~\mathrm{Sign}(n_1))!}{(n ~ \mathrm{Sign}(n))!}} \mathcal{L}_{n_1}^{n-n_1}(l_B^2\mathbf{q}^2/2)
\end{equation}
where $\mathcal{L}$ is the generalized Laguerre polynomial.

Eluding several steps, one can show that the bare polarization function is 

\begin{equation}
    \Pi^{0}(q,\omega) =\sum_{n=1}^{N_{F}}\Pi_n^{\lambda_F}(q,\omega) +\Pi_{vac}(q,\omega)
\end{equation}
where the sum runs up to the last occupied Landau level denoted by the index $N_{F}$, and the last term corresponds to the vacuum polarization for undoped graphene.

The polarizability $\Pi_n^{\lambda_F}(q,\omega)$ associated to each Landau level can be written as a sum over all the Landau levels up to an ultraviolet cutoff $N_c$ , where the weight of each level in the sum reads : 

\begin{equation}
    \Pi_{nn'}^{\lambda \lambda'}(q,\omega)= \frac{F_{nn'}^{\lambda \lambda'}(q)}{\lambda \epsilon_n -\lambda' \epsilon_{n'}+\omega +i\delta} + (\omega^{+} \rightarrow \omega^{-})
\end{equation}
where $F$ is directly related to the matrix element $\langle n | e^{i\mathbf{q}  \mathbf{r}}| n' \rangle$, and $\delta$ accounts for disorder-induced Landau level broadening. In all the calculations presented in this paper, we set $\delta = 0.05 v_{F}/l_{B}$, corresponding to a clean sample.

\begin{figure}[h!]
\centerline{\includegraphics[width=14cm]{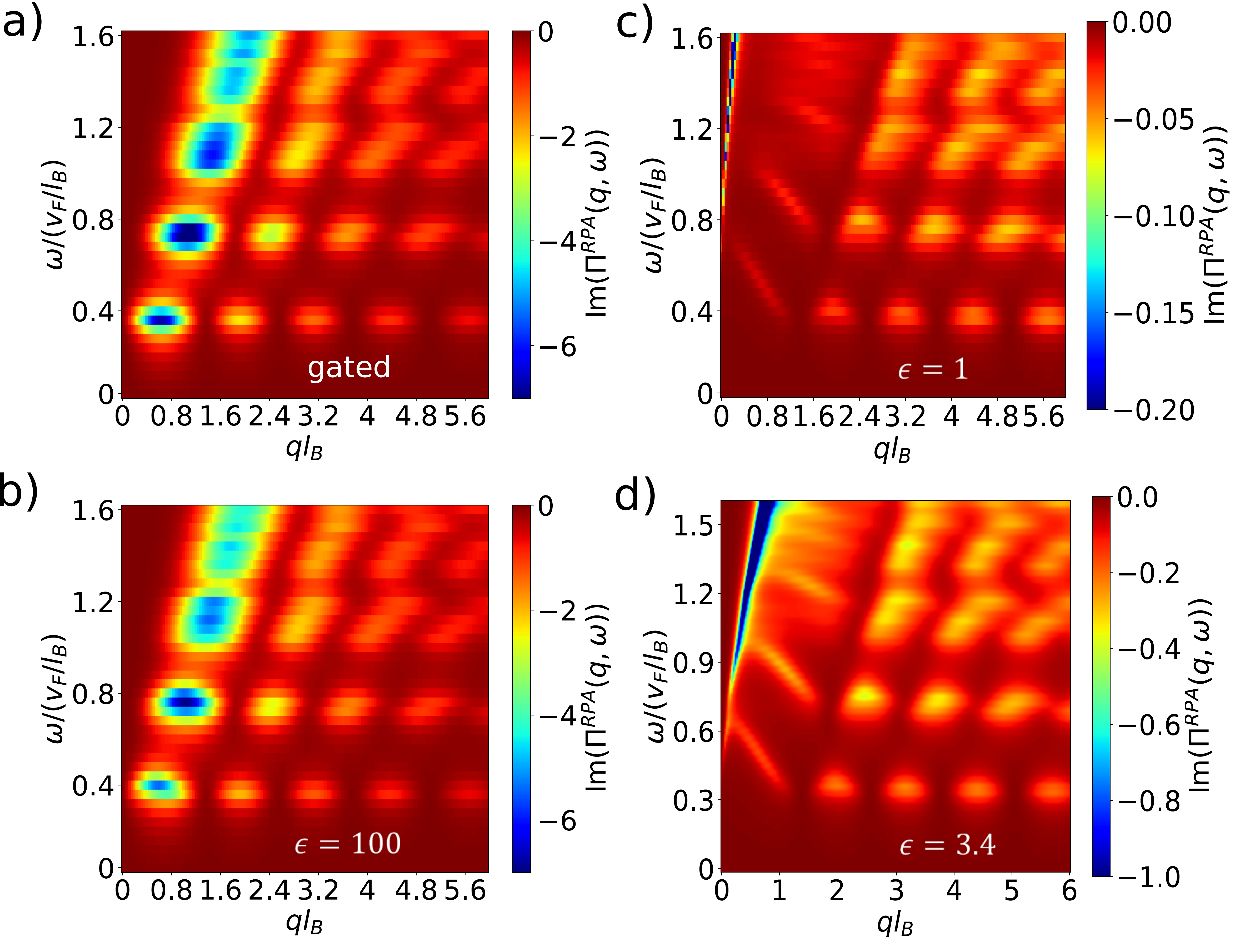}}
\caption{RPA calculations of the MLG renormalized polarization function $\Pi^{RPA}$ for $N_{F}=7$ and $B=2~\mathrm{T}$, and role of gate screening and interactions. Panel a) shows the case of hBN encapsulation with an Au gate mimicking the AuS2 device (Eq. 2 , with $d_1=32~\mathrm{nm}$, $d_2=60~\mathrm{nm}$, and dielectric functions for hBN $\epsilon_x$ and $\epsilon_z$ described in Ref. [\onlinecite{Principi2017prl}]). The PHES is very similar to that of an interaction-less ($\epsilon_b=100$ in Eq. \ref{potFAKE}) ungated sample shown in  panel b). Panels c) and d) focus on the role of interactions in the ungated case (Eq. \ref{potFAKE}), with $\epsilon_b=1$ (panel c) and $\epsilon_b=3.4$ (panel d). }
 \label{ComparMLG_Pi}
\end{figure}

This bare polarization is converted into the renormalized polarization function $\Pi^{RPA}$ using Eq.(\ref{EqRPA}). Before looking at the magneto-optical conductivity spectrum that is important for our breakdown impedance criterion, it is worth directly looking at the polarization function for the spectral weight of the collective excitations, that are represented in Fig.\ref{ComparMLG_Pi} for the same parameters $N_{F}=7$ and $B=2~\mathrm{T}$ in four cases : i) considering the presence of a gate with a screened potential (Eq. 2 with the same geometry as AuS2 sample, i.e. $d_1=32~\mathrm{nm}$, $d_2=60~\mathrm{nm}$, and dielectric functions for hBN $\epsilon_x$ and $\epsilon_z$ described in Ref. [\onlinecite{Principi2017prl}] ) , ii) in the absence of a gate (Eq. 1) with $\epsilon_b$=1, iii) in the absence of a gate (Eq. 1) with $\epsilon_b=100$ to suppress interactions and iv) in the absence of a gate (Eq. 1) with an intermediate value $\epsilon_b=3.4$.  

\begin{figure}[h!]
\centerline{\includegraphics[width=14cm]{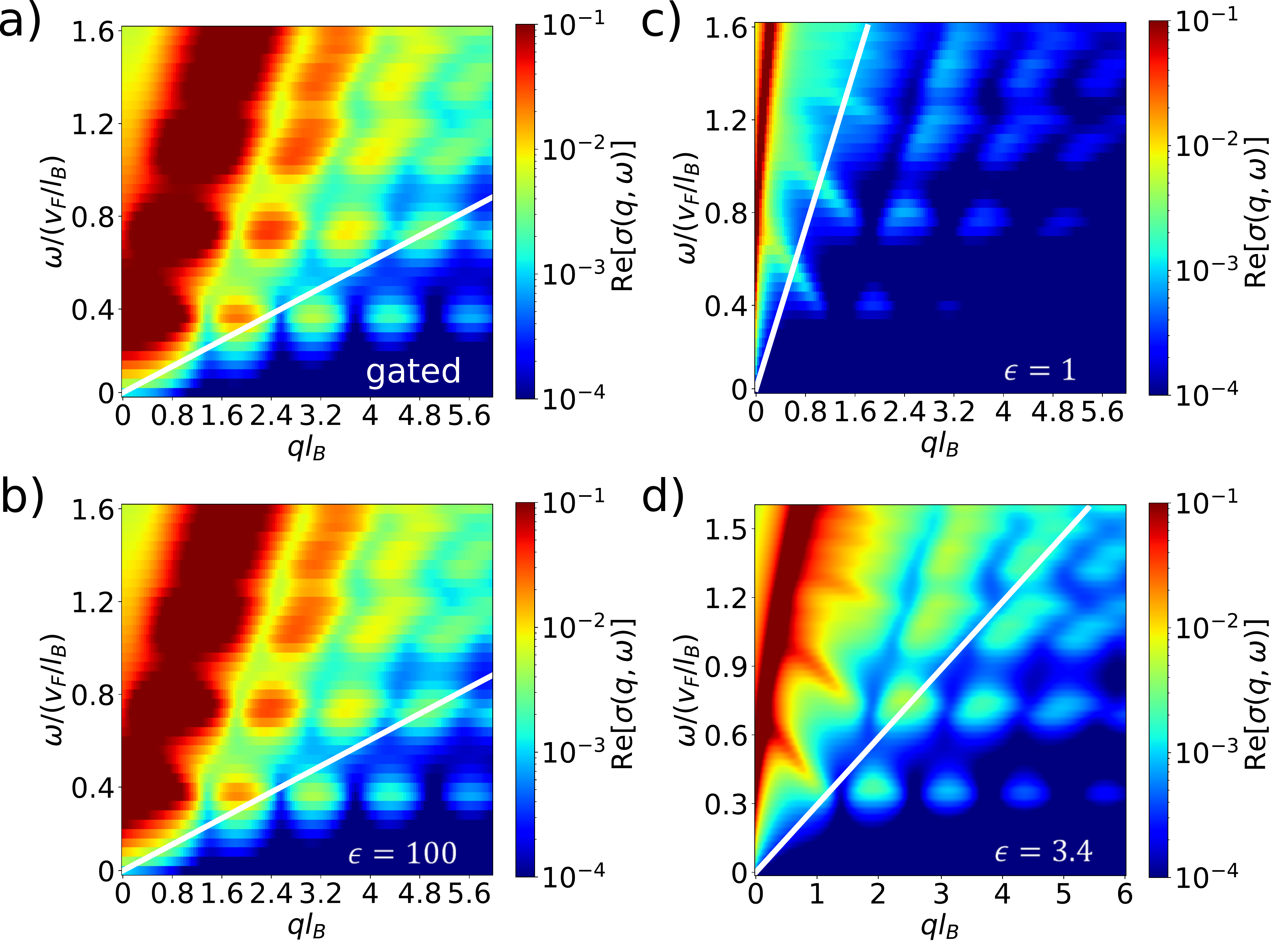}}
\caption{RPA calculations of the MLG magneto-optical conductivity $\sigma_{MO}$ for $N_{F}=7$ and $B=2~\mathrm{T}$, and role of gate screening and interactions. The conductivity values are represented in units of the Hall conductance $N_FG_K$. Panel a) shows the case of hBN encapsulation with an Au gate mimicking the AuS2 device (Eq. 2 , with $d_1=32~\mathrm{nm}$, $d_2=60~\mathrm{nm}$, and dielectric functions for hBN $\epsilon_x$ and $\epsilon_z$ described in Ref. [\onlinecite{Principi2017prl}]). The conductivity PHES is very similar to that of an interaction-less ($\epsilon_b=100$ in Eq. \ref{potFAKE}) ungated sample shown in  panel b). Panels c) and d) focus on the role of interactions in the ungated case (Eq. \ref{potFAKE}), with $\epsilon_b=1$ (panel c) and $\epsilon_b=3.4$ (panel d). The white line in panel a), b) corresponds to the measured breakdown velocity for the AuS2 device $\omega=v_{bd}q$ with $v_{bd}=0.14v_F$. The lines in panel c) and d) correspond to higher breakdown velocities following the discussion in the main text. In all cases, the breakdown Doppler line separates the high-conductivity and low-conductivity PHES sectors at $\sigma_{MO}\sim\;10^{-2}\;N_FG_K$.    }
 \label{ComparMLG}
\end{figure}

As mentioned in the main text, these calculations show the presence of a spectral gap, but they are not sufficient to explain the QHE breakdown. Although we will be mostly interested in the magneto-exciton sector of the excitations spectra for our breakdown study, it is worth mentioning that we do recover the expected behavior for the magneto-plasmon excitation (low-q sector of the colorplots): in the screened case (panels a and b), the dispersion $\omega(q)$ is linear with $v_{MP} \simeq v_F$, whereas we observe the typical $\omega \propto \sqrt{q}$ for unscreened plasmons in panels c) and d). The representation of magneto-optical conductivity deduced from $\Pi^{RPA}$ is plotted in Figure \ref{ComparMLG} for the same four configurations.

As can be seen from this Figure, the magneto-excitons spectral weight is severely suppressed by the presence of interactions (see panel c where the interactions are maximized, and panel d where the interactions are reduced), but the presence of a close local gate, that restricts the Coulombian interaction at shorter range, restores the magneto-optical conductivity spectrum observed in the non-interacting case (panel a). Note in passing that the spectrum presented in panel a), obtained for $N_{F}=7$, is in agreement with the analysis presented in the main text for $N_{F}=3$ and $N_{F}=12$, and confirms the doping-independent value of the breakdown velocity in monolayer graphene.
As explained in the main text, we interpret the constant value of the breakdown velocity in MLG as stemming from a criterion on the value of the magneto-optical conductivity $\sigma_{MO}(q,\omega)/N_F G_K \sim 10^{-2}$ (corresponding to yellowish regions on the different colorplots). The different graphene environments plotted in Figure \ref{ComparMLG} clearly indicate that the MLG breakdown velocity depends on the strength of the interactions, as exemplified by the comparison between panels c) (corresponding to strong interactions) , panel d) (corresponding to intermediate interactions) and panels a-b) where interactions are suppressed ; we would indeed observe different breakdown velocities in these three cases in order to reach the same criterion on the value of  $\sigma_{MO}(q,\omega)/N_F G_K \sim 10^{-2}$, which is highlighted by the white lines in each panel of Figure \ref{ComparMLG}.

\subsection{Bilayer graphene}
We rely once again on the derivation presented in Ref. [\onlinecite{Roldan2009prb}] for the polarization function of a 2DEG (that approximates the one of a BLG sample). 

In the case of a 2DEG, the Hamiltonian is simply the Hamiltonian of free electrons in a magnetic field $\mathcal{H}=\frac{\mathbf{\Pi}^2}{2m}$ where $\mathbf{\Pi}=\mathbf{p}+e\mathbf{A(r)}$ takes into account the coupling to the magnetic field $\mathbf{B} = \mathrm{\mathbf{rot}} \mathbf{A(r)}$. One obtains straightforwardly the usual Landau level spectrum with eigen-energies $\epsilon_n=\hbar \omega_c (n+\frac{1}{2})$ where $\omega_c=eB/m$ is the cyclotron frequency. 
\\From the calculation of the Green's function with the help of the usual eigenstates $|n \rangle$ , we can obtain the polarization function for the 2DEG :

\begin{equation}
    \Pi^0 (q,\omega)= \sum_{n=0}^{N_F} \sum_{n'=N_F +1}^{\infty} \frac{F_{nn'}(q)}{(n-n')\omega_c +\omega +i \delta} + ( \omega^{+} \rightarrow \omega^{-})
\end{equation}
where the form factor $F_{nn'}(q)$ is defined in terms of a generalized Laguerre polynomial : 

\begin{equation}
    F_{nn'}(\mathbf{q})=e^{-l_B^2 q^2 /2}\biggl(\frac{l_B^2 q^2}{2} \biggr)^{n_{>}-n_{<}} \frac{n_{<}!}{n_{>}!} \biggl[\mathcal{L}_{n_{<}}^{n_{>}-n_{<}}(\frac{l_B^2 q^2}{2}) \biggr]^{2}
\end{equation}
with $n_{>}=\mathrm{max}(n,n')$ and $n_{<}=\mathrm{min}(n,n')$.

  \begin{figure}[h!]
\centerline{\includegraphics[width=14cm]{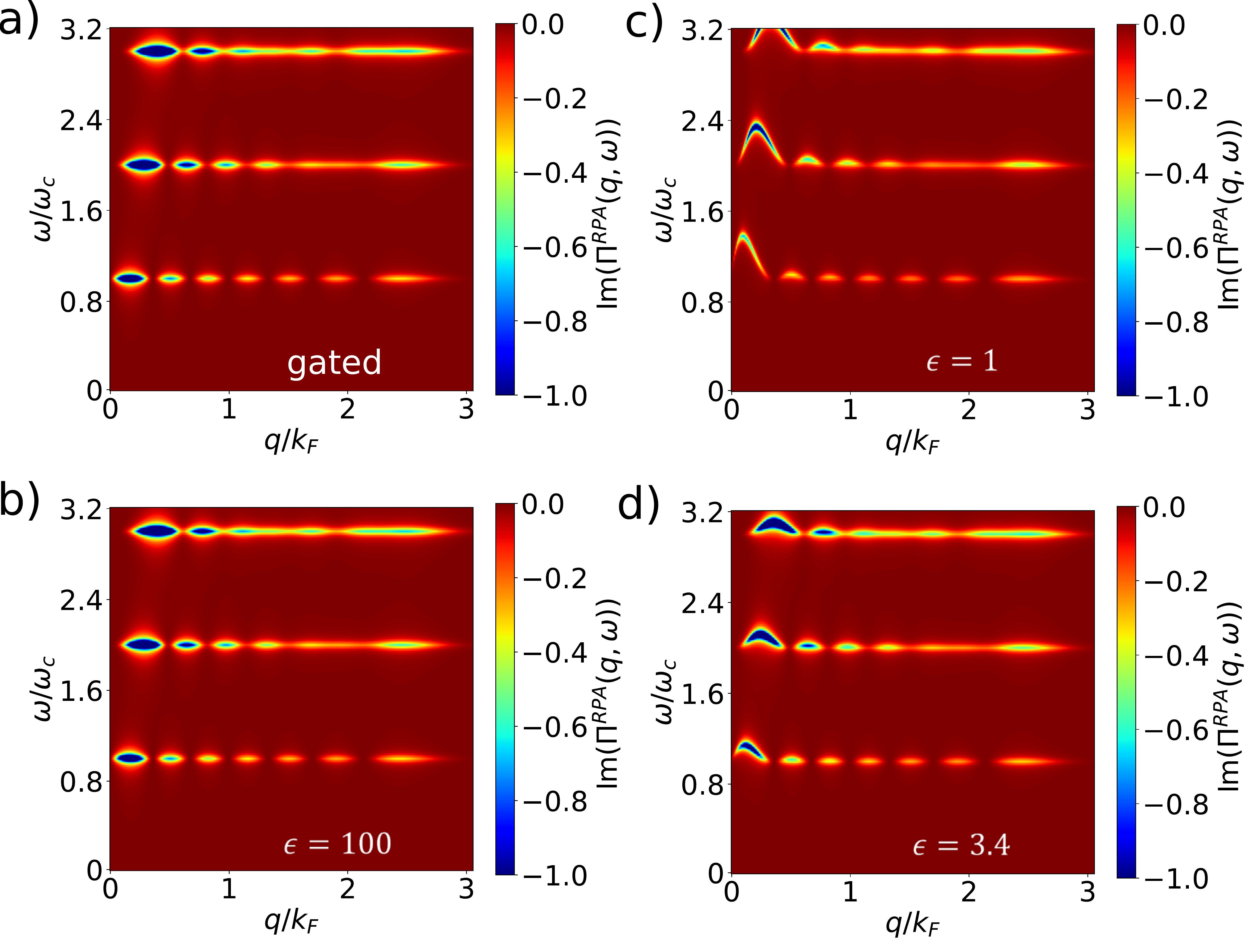}}
\caption{RPA calculations of the BLG renormalized polarization function $\Pi^{RPA}$ for $N_{F}=7$ and $B=2~\mathrm{T}$. Panel a) shows the case of hBN encapsulation with an Au gate (Eq. 2 , with $d_1=32~\mathrm{nm}$, $d_2=60~\mathrm{nm}$, and dielectric functions for hBN $\epsilon_x$ and $\epsilon_z$ described in Ref. [\onlinecite{Principi2017prl}]), which is very similar to the interaction-less unscreened case ($\epsilon_b=100$ in Eq. \ref{potFAKE}) displayed in panel b). Panels c) and d) focus on the effect of the interactions in the unscreened case (Eq. 1), with $\epsilon_b=1$ (c) and $\epsilon_b=3.4$ (d) respectively.}
 \label{ComparBLG_Pi}
\end{figure}

From this bare polarization we get the real response function taking into account the interactions within the RPA. In a first step we can look at the polarization function spectra defining the collective modes in the same four cases as for monolayer graphene in previous section. We keep the same parameters $N_{F}=7$ and $B=2~\mathrm{T}$ : i) considering the presence of a gate with a screened potential (Eq. 2 with the same geometry as AuS2 sample, i.e. $d_1=32~\mathrm{nm}$, $d_2=60~\mathrm{nm}$, and dielectric functions for hBN $\epsilon_x$ and $\epsilon_z$ described in Ref. [\onlinecite{Principi2017prl}] ), ii) in the absence of a gate (Eq. 1) with $\epsilon_b$=100 to suppress interactions, iii) in the absence of a gate (Eq. 1) with $\epsilon_b=1$ to maximize the effect of interactions and iv) in the absence of a gate (Eq. 1) with an intermediate $\epsilon_b=3.4$. The result is plotted in Figure \ref{ComparBLG_Pi}.

\begin{figure}[h!]
\centerline{\includegraphics[width=14cm]{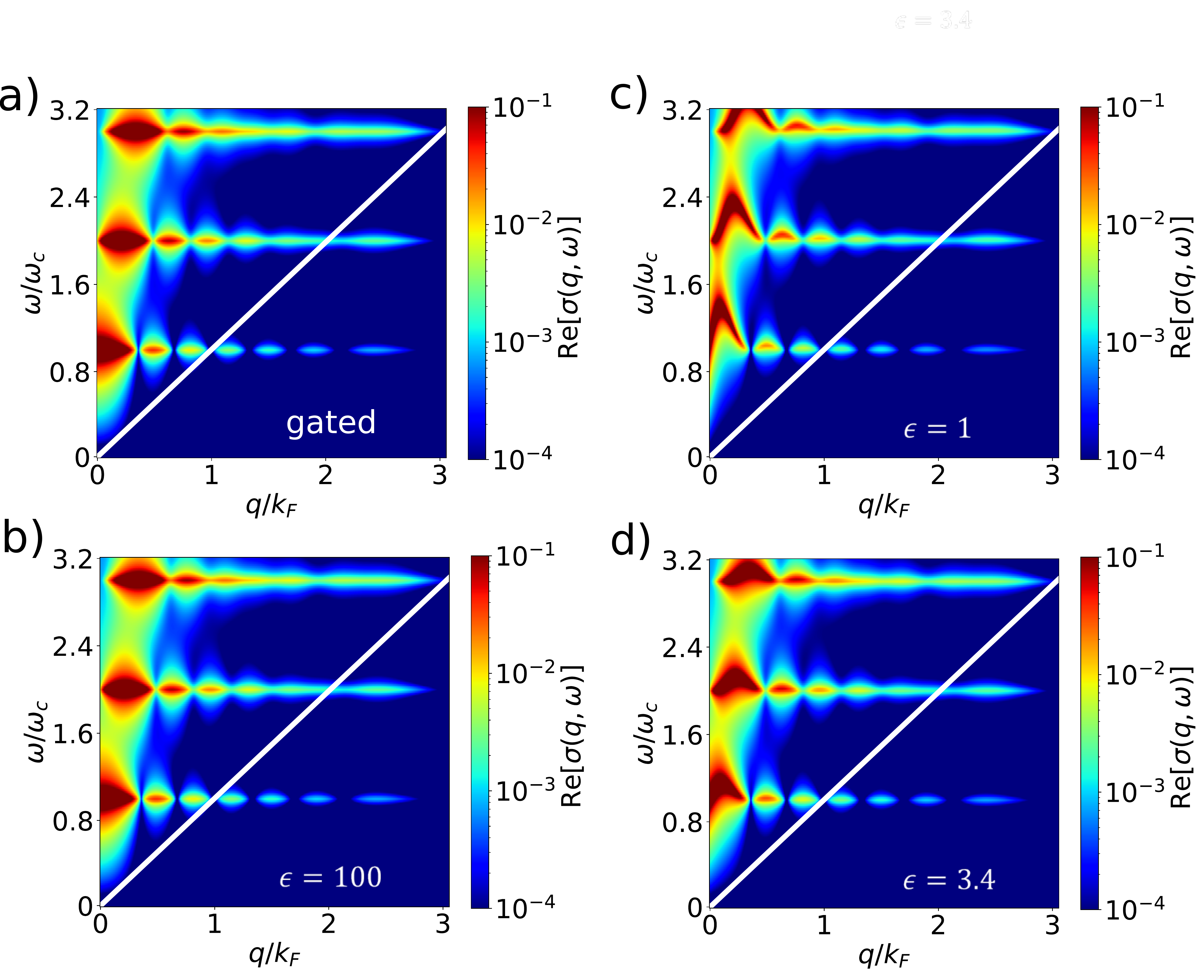}}
\caption{RPA calculations of the BLG magneto-optical conductivity $\sigma_{MO}$ for $N_{F}=7$ and $B=2~\mathrm{T}$. The conductivity values are represented in units of the Hall conductance $N_FG_K$. Panel a) shows the case of hBN encapsulation with an Au gate (Eq. 2 , with $d_1=32~\mathrm{nm}$, $d_2=60~\mathrm{nm}$, and dielectric functions for hBN $\epsilon_x$ and $\epsilon_z$ described in Ref. [\onlinecite{Principi2017prl}]), which is very similar to the interaction-less unscreened case (Eq. \ref{potFAKE}) with $\epsilon_b=100$ displayed in panel b). Panels c) and d) focus on the effect of the interactions in the unscreened case (Eq. 1), with $\epsilon_b=1$ (c) and $\epsilon_b=3.4$ (d) respectively. The white line corresponds to the breakdown velocity described in Ref.[\onlinecite{Yang2018prl}].}
 \label{ComparBLG}
\end{figure}

We recover the expected features of the PHES, with a spectral gap corresponding to the cyclotron frequency $\omega_c$ for BLG, as well as the magneto-exciton branches. It appears already at this point that the effect of interactions is very tiny. This is confirmed by the computation of the magneto-optical conductivity for a bilayer graphene, displayed in Figure \ref{ComparBLG}.

 As opposed to the MLG case, the role of interactions in the magneto-excitons optical conductivity is much less pronounced, with zero to little variation of the spectral weight for the BLG magneto-excitons when tuning the interactions. As explained in the main text, the existence of an impedance matching criterion on the magneto-exciton conductivity sheds a new light on the fortuitous agreement between the BLG breakdown velocity and the Zener critical velocity associated to inter-Landau-level tunneling.